\documentclass{article}
\usepackage{spconf,amsmath,graphicx}
\usepackage{algorithm}
\usepackage[noend]{algpseudocode}
\usepackage{multirow}
\usepackage{hyperref}
\usepackage{balance}

\title{MANNER OF ARTICULATION DETECTION USING CONNECTIONIST TEMPORAL CLASSIFICATION TO IMPROVE AUTOMATIC SPEECH RECOGNITION PERFORMANCE}
\name{Pradeep R and Sreenivasa Rao K\thanks{We thank Tata Consultancy Services (TCS) for sponsoring the research under TCS-Research Scholar Programme.}}
\address{Department of Computer Science and Engineering,\\ Indian Institute of Technology, Kharagpur, India \\ {\small \tt \{pradeep\_raj31, ksrao\}@iitkgp.ac.in}
}
\begin{document}
%
\maketitle
\begin{abstract}

Conventionally, the manner of articulations in speech signal are derived using  discriminative signal processing techniques or deep learning approaches. However, training such complex systems involves feature extraction, phoneme force alignment and deep neural network training. In our work, we initially detect the manner of articulations without phoneme alignment using an end-to-end manner of articulation modeling based on connectionist temporal classification (CTC). The manner of articulation knowledge is deployed in the conventional character CTC path to regenerate the new character CTC path. The modified manner based character CTC is evaluated on open source speech datasets such as AN4, LibriSpeech and TEDLIUM-2 and it outperforms over the baseline character CTC.

\end{abstract}
\section{Introduction}
\label{sec:intro}

Automatic speech recognition (ASR) has emerged as one of the active areas of technology with a lot of research and development efforts mainly in speech processing communities. The conventional acoustic models (AMs) of an ASR system followed a generative approach based on Hidden Markov Models (HMMs) \cite{hinton2012deep} where the emission probabilities of each state were modeled with a Gaussian Mixture Model (GMM). Later, the performance of ASR has been improved dramatically by the introduction of deep neural networks (DNNs) as acoustic models \cite{dahl2012context} \cite{maas2017building}. In the hybrid HMM/DNN approach, DNNs are used to classify speech frames into clustered context-dependent (CD) states (i.e., senones). On a variety of ASR tasks, DNN models have shown significant gains over the GMM models. However, the HMM-GMM or HMM-DNN pipelines are highly complex as it involves multiple training strategies (such as CI phones, CD senones), various parameters (such as dictionaries, decision trees) and the performance depends on optimal choice of number of senones, Gaussians, etc.

More recent work has been focused on solutions which involves a simple paradigm which come close to end-to-end systems. 
 On this aspect, Graves et al. \cite{graves2006connectionist} introduced the connectionist temporal classification (CTC) objective function to infer speech-label alignments automatically without having to rely on an alignment between the audio sequence and the symbol sequence. This CTC technique is further investigated in \cite{graves2014towards} \cite{sak2015learning} on large-scale acoustic modeling tasks. On the other hand, the attention-based method \cite{bahdanau2016end} exploits an attention mechanism to perform alignment between acoustic features and recognized symbols. However, the basic temporal attention mechanism is too flexible in the sense that it allows extremely non-sequential alignments.
This is rational for applications such as machine translation where input and output word order are different \cite{bahdanau2014neural}. However in phone attribute detection, the acoustic features and the corresponding outputs proceed in a monotonic way. Since CTC permits an efficient computation of a strictly monotonic alignment using dynamic programming, we propose to train a CTC-based manner of articulation detector to detect vowels, semi-vowels, nasals, fricatives and stop consonants.

Speech attributes were detected using discriminative features at the front-end and training the classifier \cite{lee2007overview}. Signal processing approaches are used for automatic and accurate detection of the closure-burst transition events of stops and affricates \cite{ananthapadmanabha2014detection}. 
Later deep learning techniques were used to detect speech attributes \cite{siniscalchi2013exploiting}.
However, training such complex systems involves feature extraction, phoneme force alignment and deep neural network training. Recently, Cernak et al., \cite{cernak2018nasal} exploited a solution to train a nasal sound detector without phone alignment using an end-to-end phone attribute modeling based on the connectionist temporal classification.

Recent success on nasal detection using CTC  motivated us to extend their framework for detecting five broad manners of articulation namely vowel, semi-vowel, nasal, fricative and stop consonant. In the first part of our work, we extend CTC based nasal and non-nasal detection framework \cite{cernak2018nasal} to detect five broad manners of articulation. Later, we propose a mechanism where the manner of articulation detection knowledge is deployed in decoding a speech utterance and study its impact in ASR performance. The main addressed question of this work was if it is possible to train a CTC manner detector without phone alignment and embed this knowledge in ASR decoding. 


\vspace{-0.5cm}
\section{Manner of Articulation Detection using CTC}
\label{sec:mannerDetect}
\begin{figure}[h]
\begin{minipage}[b]{1.0\linewidth}
  \centering
  \centerline{\includegraphics[width=5.0cm,height=1.0cm]{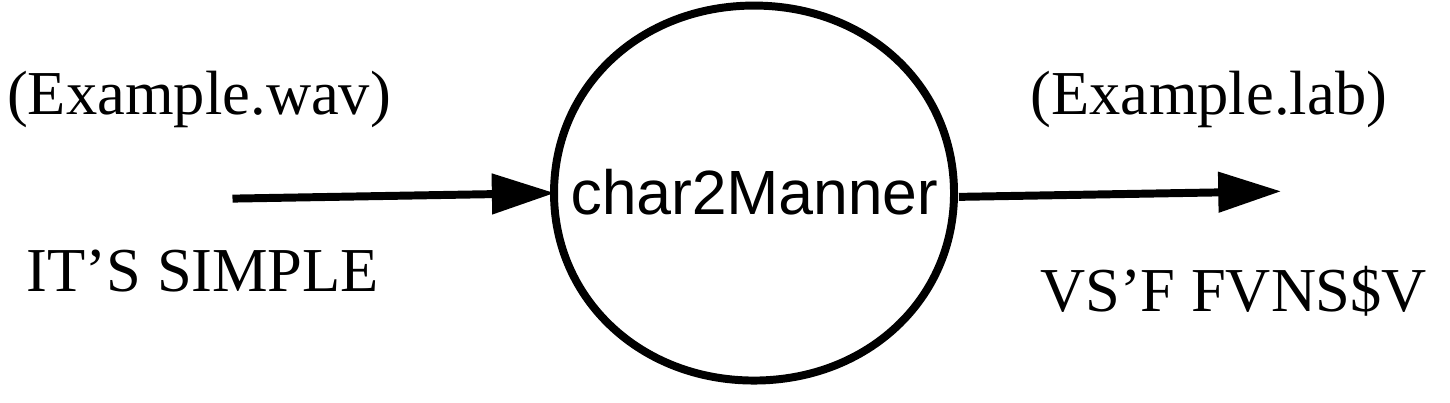}}
  \caption{\it Character to Manner of Articulation Text Transcript}
  \label{fig:char2mannerInit}
\end{minipage}
\end{figure}

The character level transcripts are mapped to five different manners of articulation \{$V, \$, N, F, S$\} that represents vowel, semi-vowel, nasal, fricative and stop consonant respectively. Figure \ref{fig:char2mannerInit} illustrates a toy example of obtaining manner of articulation labels from the character labels.
The connectionist temporal classification (CTC) approach \cite{graves2014towards} is an objective function that allows an end-to-end training without requiring any frame-level alignment between the input and target labels.  
It introduces an intermediate representation called the CTC path. The label sequence can be represented by a set of all the possible CTC paths that are mapped to it.

For an input sequence $\textbf{X} = (x_1 , ..., x_T )$, the conditional probability $P(z|\textbf{X})$ is then obtained by summing over all the probabilities of all the paths that corresponding to the target label sequence y after inserting the repetitions of labels and the blank tokens, i.e.,
\begin{eqnarray}
P(z|X) = \sum_{y'\in \phi (z) }P(y'|X) = \sum_{y'\in \phi (z) }\prod_{t=1}^{T}P(y_t'|x_t)
\end{eqnarray}
where $\phi(z)$ denotes the set of all possible paths that correspond to $z$ after repetitions of labels and insertions of the blank token. The conditional probability of the labels at each time step, $P(y_t^{'}|x_t)$, is generally estimated using a RNN (LSTM/GRU). The model can be trained to maximize Equation 1 by using gradient descent, where the required gradients can be computed using the forward-backward algorithm \cite{graves2006connectionist}.

In order to detect manners of articulation using CTC, the softmax output nodes are set to $M+k$ where $M$ is is the number of manners and $k$ characters are added to include blank ($<$) or space ($>$). The manner of articulation detector using CTC is shown in left part of Figure \ref{fig:mannerplusCharCTC}. The bottom of the network is two layers of convolutions over both time and frequency domains. Temporal convolution is commonly used in speech processing to efficiently model temporal invariance for variable length utterances. Convolution in frequency attempts to model spectral variance due to speaker variability and it has been shown to further improve the performance \cite{sainath2013deep}. Following the convolutional layers are bidirectional recurrent layers.
After the bidirectional recurrent layers, a fully connected layer is applied and the output is produced through a softmax function computing a probability distribution over the target labels {blank, vowel, semi-vowel, nasal, fricative, stops, space}. The model is trained using the CTC loss function. To accelerate the training procedure, Batch Normalization \cite{ioffe2015batch} is applied on hidden layers.

\vspace{-0.5cm}
\section{Character CTC derived from Manner of Articulation CTC}
\label{sec:charCTCfromManner}
\begin{figure}[htb]
\begin{minipage}[b]{1.0\linewidth}
  \centering
  \centerline{\includegraphics[width=8.0cm,height=5.0cm]{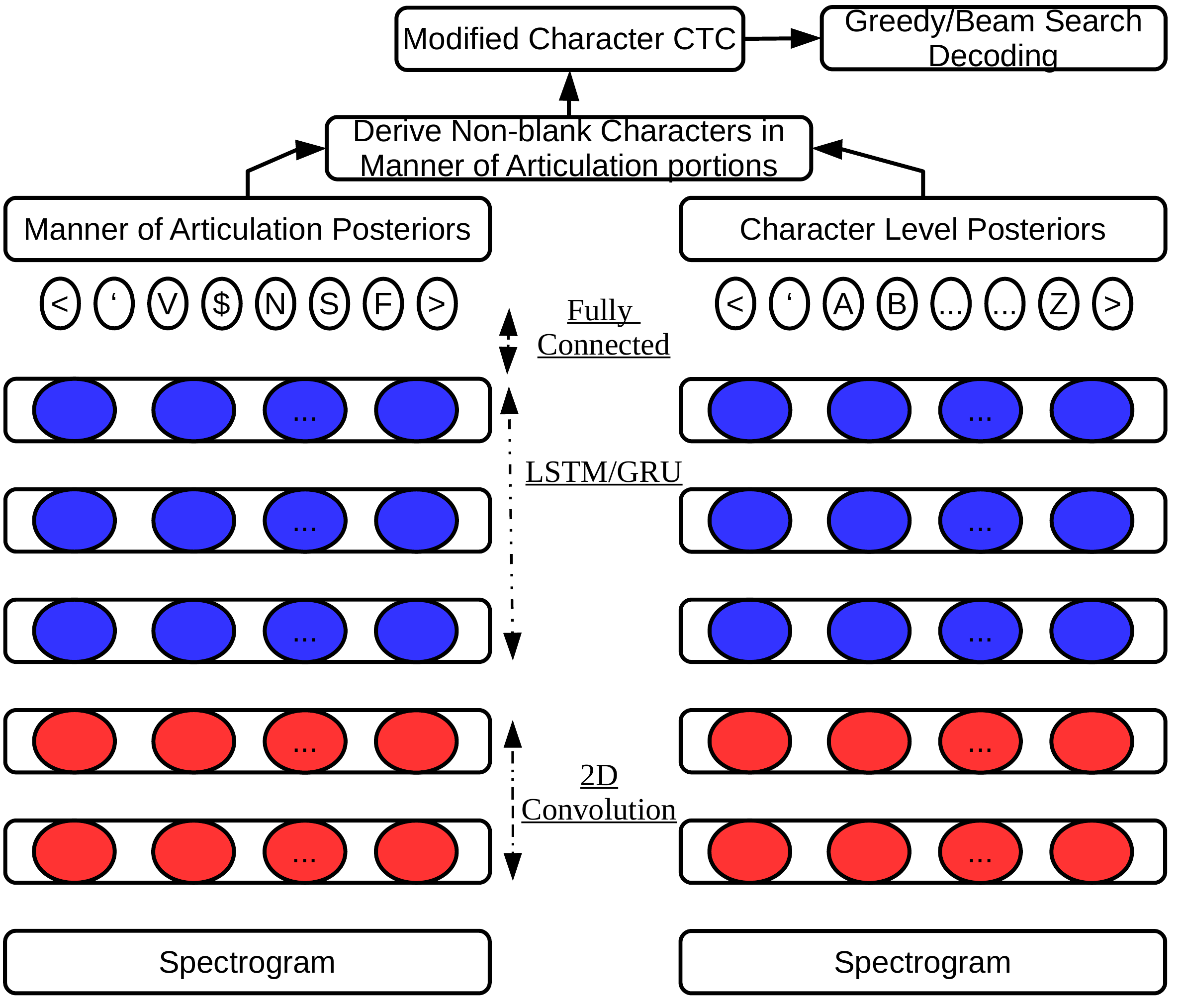}}
  \caption{\it Character CTC derived from manner CTC}
  \label{fig:mannerplusCharCTC}
\end{minipage}
\end{figure}

We separately train manner of articulation CTC detector and the conventional character based CTC detector. Figure \ref{fig:mannerplusCharCTC} shows the overview of obtaining character CTC by combing the manner of articulation posteriors and the baseline character posteriors. 

\vspace{-0.8cm}

\subsection{Modified character CTC}

The basic version of obtaining manner based character CTC is shown in Algorithm 1. The modified character index is initialized with blank labels (line 4) and the previous most probable character index is initialized with space index (5).
\makeatletter
\def\BState{\State\hskip-\ALG@thistlm}
\makeatother
\begin{algorithm}
\caption{: $Result \gets MannerbasedCharCTC(Data)$}\label{euclid}
\begin{algorithmic}[1]
\State ${\textbf{Data}: postManner, postChar, labelsChar}$
\State ${\textbf{Result}: modChStr}$
\State $blankInx \gets 1 $
\State $modCharInx \gets ones(1,length(postC)) $
\State $prevInx \gets find(labelsChar==>); spaceInx$
\State $mannerInx \gets argmax(postM)$
\State $[nBInx,nBseg] \gets find(mannerInx\neq1)$
\For {$i = 1 : length(nBInx)$} :
\State $Ch(i,:) \gets argmax([0,postC([2:end],nBSeg)])$
\State $charList \gets sort_f(Ch \not\in prevInx)$
\State $currInx \gets charList[1]$
\State $modCharInx(nBSeg) \gets currInx$
\State $prevInx \gets currInx$
\EndFor
\State $modChStr \gets sym2Char(modCharInx,labelsChar)$
\State $\textbf{return}$  $modChStr$
\end{algorithmic}
\end{algorithm}
 Then, we find the frame level index by calculating the most probable manner of articulation portions obtained from posteriors manner, $(postManner)$ (6). We find the non-blank index,$nBInx$ in the manner CTC and their start and end frame durations, $nBSeg$ (7). 
 The algorithm iterates over all non-blank portions obtained from the manner CTC. We iteratively find the index of the most probable character segment by forcing the blank probabilities to zero (9). This is to ensure that the non-blank segment derived from manner CTC always emits a non-blank character. Later, in the non-blank manner segments, we find the most probable characters and are sorted according to the frequency of occurrence ($sort_f$) within the manner alignments (10). 
 In the next phase, we find the characters excluding the previous index so that a new symbol is generated. We compare the current character generated for a manner peak with the previous character index and is ensured that the current character is always new as compared to the previous character (11-12). The current index now acts as a previous index and the process is repeated for all frames (13).  After the last time-step, the best character indices are converted to character symbols and is returned as a result (14-15).
 The basic idea of the proposed method is to observe the non-blank peaks in the manner CTC and it is forced to emit a non-blank symbol thereby some of the characters missed out from blank dominating probabilities are nullified.



\vspace{-0.6cm}
\subsection{Illustrative Example}

\begin{figure}[htb]
\begin{minipage}[b]{1.0\linewidth}
  \centering
  \centerline{\includegraphics[width=8.0cm,height=4.5cm]{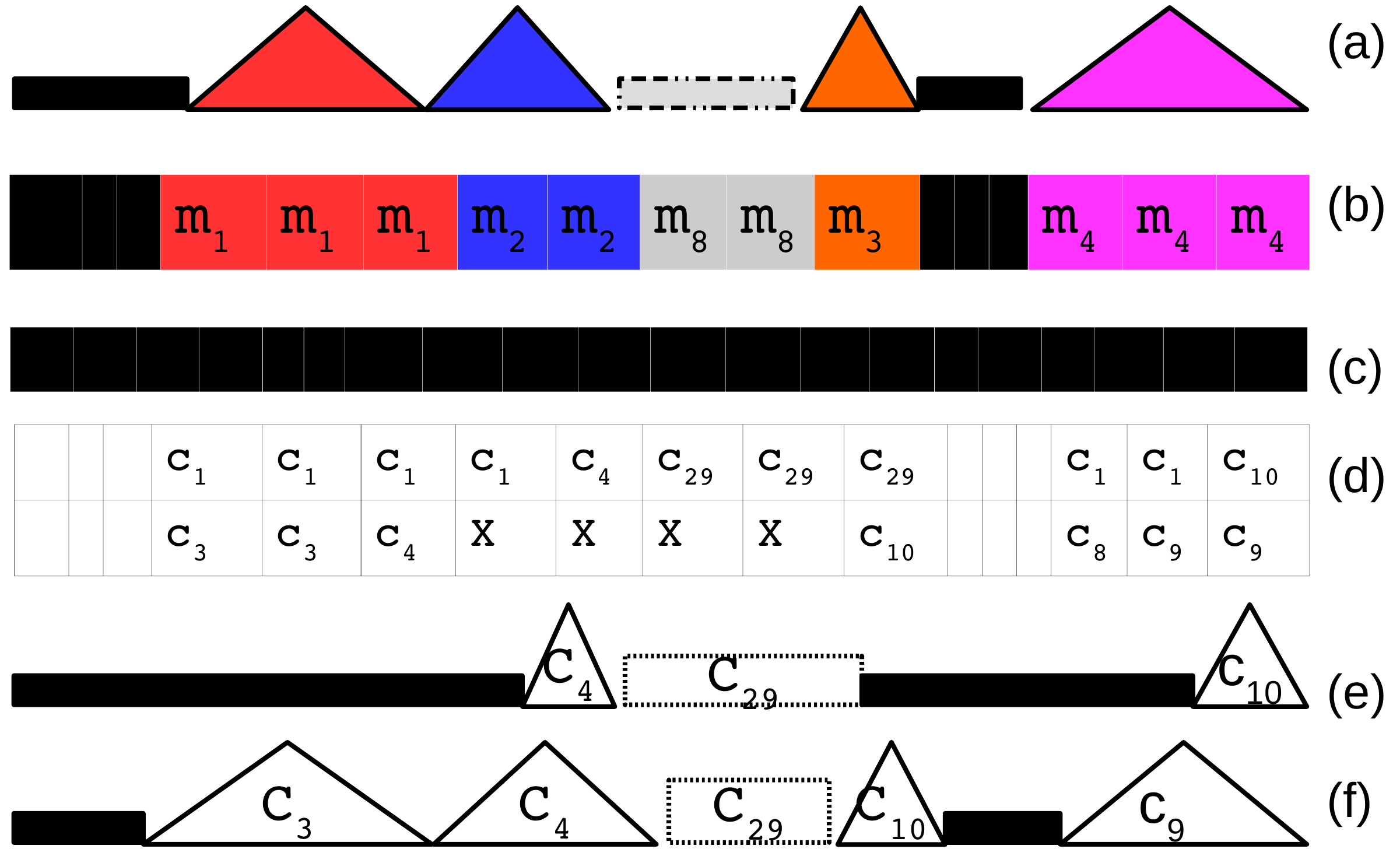}}
  \caption{\it (a) Manner of articulation predicted labels using manner CTC  detector
(b) Most probable manner index at frame level
(c) New character index initialization to blank symbols
(d) Sorted character indices obtained using character CTC detector for each frame
(e) Most probable character index using state-of-the art character CTC detector
(f) Character indices derived from manner of articulation CTC detector}
  \label{fig:CharViaMannerCTC}
  \vspace{-0.4cm}
\end{minipage}
\end{figure}

Figure \ref{fig:CharViaMannerCTC} shows an example of obtaining the character indices based on manner CTC detector. Initially the manner of articulation predicted labels for a test utterance are found using manner CTC detector.
 Different colors indicate different manners of articulation. We denote the most probably predicted blank symbols in black color and space character by dotted rectangle for illustration. We consider only those portions where the manner CTC detector emits non-blank symbols. The most probable manner index at frame level is shown in Figure \ref{fig:CharViaMannerCTC}(b). We initialize the manner based CTC characters to blank symbols as shown in Figure \ref{fig:CharViaMannerCTC}(c). Figure \ref{fig:CharViaMannerCTC}(d) shows the sorted list of character indices  for each frame based on character posterior probabilities. They are generated using baseline character CTC detector.  We denote the characters $C_1$ and $C_{29}$ for representing blank and space characters respectively for illustration and $C_2$-$C_{28}$ for representing non-blank characters.  The greedy search on these character probabilities picks up the most probable character index at each frame and it is represented in Figure \ref{fig:CharViaMannerCTC} (e). The decoded sequence generated using best path decoding or greedy search is $C_4 C_{29} C_{10}$.

We find the start and end frame index of manner CTC peak.
If the manner CTC emits  a non-blank symbol, then the character CTC is forced to emit a non-blank character. This can be achieved by iteratively finding the most probable element in decreasing order of the posteriors probabilities and finding the optimal character by sorting the characters according to their frequency of occurrence in each segment.
For instance $m_1$ is the most probable manner index and the corresponding character index is $c_1$ which is a blank symbol. So we find the second maximum character indices and sort them according to their frequency of occurrence and hence emit a single non-blank symbol. We will also compare the current index emitted with the previous index and if both are same then we emit different index by choosing appropriate symbol.
The decoded sequence generated using proposed manner based character CTC is $C_3 C_4 C_{29} C_{10} C_9$.

\vspace{-0.4cm}

\section{Experiments}
\label{sec:Experiments}
\vspace{-0.5cm}
\subsection{Data}
We used three open source databases for training the manner and character CTC systems:
(1) AN4 \footnote{http://www.speech.cs.cmu.edu/databases/an4/} \-- the database contains alpha numeric speech data having 948 training and 130 test utterances. The dataset provides a good sample to achieve deterministic results to scale up with larger datasets.
(2) LibriSpeech \footnote{http://www.openslr.org/resources/12/} \-- the data are sampled at 16 kHz, and the training part of the corpus are with size approximately 100, 360 and 490 hours respectively. In our experiments, we use 100 hours train-clean corpus.
(3) TEDLIUM-2 \footnote{http://www.openslr.org/7/} \-- the English-language TED talks, with transcriptions, sampled at 16kHz. It contains about 118 hours of speech.

\vspace{-0.3cm}
\subsection{Training}

The training phase is based on the open source DeepSpeech \footnote{https://github.com/SeanNaren/deepspeech.pytorch} network \cite{amodei2016deep}, trained with the CTC activation function.
  The manner detector using CTC starts with two layers of 2D convolutions over both time and frequency domains with 32 channels, 41 $\times$ 11, 21 $\times$ 11 filter dimensions, and 2 $\times$ 2, 2 $\times$ 1 stride. Four next bidirectional gated recurrent layers with 400 hidden units are followed by one fully connected linear layer with 7 softmax outputs \{$blank,', vowel, semi-vowel, nasal, fricative , stop, space$\}. The GRU models have around 4.1 millions (M) parameters. The input sequence are values of spectrogram slices, 20 ms long, computed from Hamming windows with 10 ms frame shifts. 
The output (target) sequence was obtained directly from the letters of the word transcription. 
  We used 50 epochs to train all the models used for further evaluation.

\vspace{-0.5cm}
\section{Results}
\label{sec:Results}
We evaluated both character and manner CTC detectors on the test-clean data set. 
The manner of articulation error  rate (MER) is used to measure the performance of the manner CTC detector. The calculation of MER is similar to that of character error rate where the manner based reference and the obtained transcripts are compared. The reference transcripts is initially changed as per Figure\ref{fig:char2mannerInit} . 
Table \ref{tab:MER} shows obtained MERs of the CTC manner of articulation detector obtained on three datasets.
\vspace{-0.5cm}
\begin{table}[h]
\centering
\caption{\it Manner of Articulation Error Rate for Manner CTC detector }
\footnotesize\setlength{\tabcolsep}{4pt}
\renewcommand{\arraystretch}{1.0}
\label{tab:MER}
\begin{tabular}{|c|c|}
\hline
\textbf{Dataset} & \textbf{\% MER} \\ \hline
AN4 & 2.8 \\ \hline
LibriSpeech & 2.7 \\ \hline
TEDLIUM-2 & 12.2 \\ \hline
\end{tabular}
\end{table}
Table \ref{tab:WERCER} shows the  word error rate (WER) and the character error rate (CER) obtained using the baseline and the proposed method.
\vspace{-0.5cm}
\begin{table}[h]
\centering
\caption{\it WER and CER obtained using baseline CTC and the proposed method on different datasets}
\footnotesize\setlength{\tabcolsep}{5pt}
\renewcommand{\arraystretch}{1.0}
\label{tab:WERCER}
\begin{tabular}{|c|c|c|c|}
\hline
\textbf{Dataset} & \textbf{Method} & \textbf{\% WER} & \textbf{\%CER} \\ \hline
\multirow{2}{*}{AN4} & Baseline & 9.3 & 3.7 \\ \cline{2-4} 
 & Proposed & 8.9 & 3.0 \\ \hline
\multirow{2}{*}{LibriSpeech} & Baseline & 11.1 & 3.3  \\ \cline{2-4} 
 & Proposed & 10.8 & 3.0 \\ \hline
\multirow{2}{*}{TEDLIUM-2} & Baseline & 34.0  & 13.1  \\ \cline{2-4} 
 & Proposed & 33.6 & 12.5  \\ \hline
\end{tabular}
\vspace{-0.3cm}
\end{table}
The pre-trained manner of articulation models and the baseline CTC models trained with AN4 dataset is made as an open source code\footnote{https://github.com/Pradeep-Rangan/Manner-of-Articulation-Detection-using-CTC}.
It is observed that the manner of articulation knowledge in modifying the CTC path has significant impact in improving the performance of ASR.

\vspace{-0.5cm}
\subsection{Discussion}

The CNN in the used model performed 2D convolution, where the first dimension is frequency and the second dimension is time. A longer stride is usually applied to speed-up training. Using the stride in the time dimension results into time compacting of the input audio, e.g., using the stride of 2 results into 2 times less frames of the output. For applications where time alignment is required, we experimented with the stride of 1 and we observed the convergence in training.
\begin{figure}[htb]
\begin{minipage}[b]{1.0\linewidth}
  \centering
  \centerline{\includegraphics[width=8.5cm,height=5.0cm]{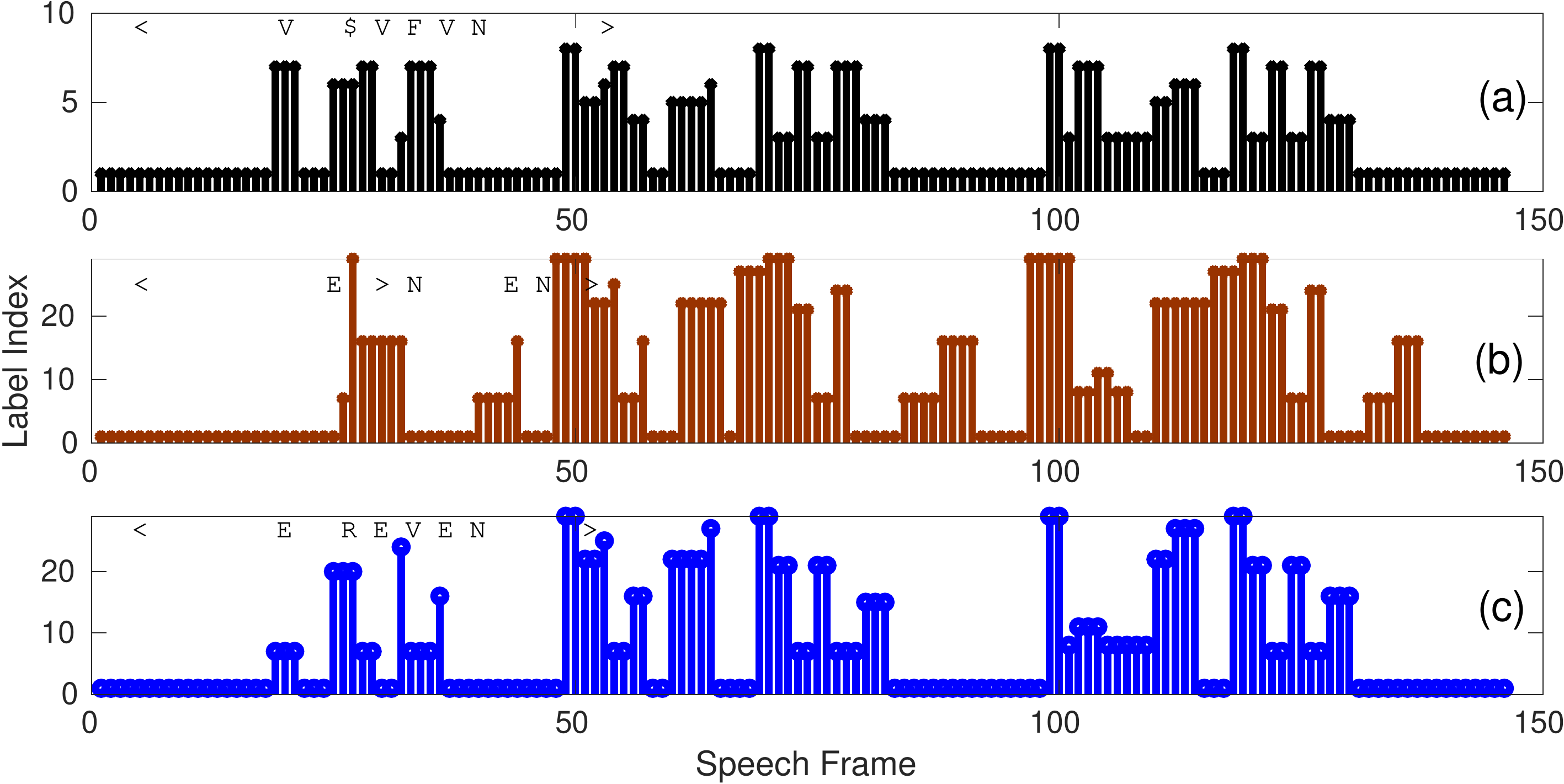}}
  \vspace{-0.5cm}
  \caption{\it (a) manner of articulation Index (b) baseline character index (c) modified manner based character index}
  \label{fig:compareResults}
\end{minipage}
\end{figure}
Figure \ref{fig:compareResults} shows an example of the label index generated on manner CTC, baseline character CTC and the modified manner based character CTC detector.
The speech utterance is from AN4 test dataset $(test/an4/wav/cen8-fcaw-b.wav)$ whose content has the sentence ``ELEVEN TWENTY SEVEN FIFTY SEVEN''. The most probable manner index is derived from the posteriors manner as shown in Figure \ref{fig:compareResults} (a). On top of the figure we illustrate some of the text transcript portions. The baseline character CTC as shown in Figure \ref{fig:compareResults} (b) generates ``E NEN TWENTY SEVEN FIFTY SEVEN'' leading to false insertion and substitution errors. The non-blank portions in the manner peaks are forced to generate non-blank characters derived from character CTC detector using proposed method. The manner of articulation based character CTC generates the optimal character labels as per the peakiness in the manner portions. Figure \ref{fig:compareResults} (c) shows the modified character index . The decoded sequence obtained using proposed method is : ``EREVEN TWENTY SEVEN FIFTY SEVEN''. It can be observed that the additional space that was generated using baseline method is nullified using the proposed method. Also the blank character propbabilties that dominated to miss out the substring `EVEN' is recovered. The generation of the character 'R' may be due to the fact that the manner of articulation has semivowel. The probabilty of occurance of L is less than that of R.


\vspace{-0.6cm}
\section{Conclusion}
\label{sec:conclude}

This paper has proposed to use the connectionist temporal classification for the end-to-end manner of articulation modeling.
The manner of articulation knowledge is deployed in the conventional character CTC path to regenerate the new character CTC path. The modified manner based character CTC is evaluated on open source speech datasets such as AN4, LibriSpeech and TEDLIUM-2 and it outperforms over the baseline character CTC.
Application of the proposed manner of articulation CTC detector in weight adaptation of baseline end-to-end ASR training is also planned for future work.

\bibliographystyle{IEEEbib}
\balance
\bibliography{strings,refs}

\end{document}